\documentclass[11pt,twoside]{article}

\usepackage{asp2006}
\usepackage{epsf}
\usepackage{psfig}
\usepackage{lscape}

\newcommand{\hii}       {H~{\sc ii}}

\begin{document}
\title{A Metallicity Map of M33}   
\author{Joshua D. Simon\altaffilmark{1} and Erik Rosolowsky\altaffilmark{2}}   
\altaffiltext{1}{Department of Astronomy, California Institute of
  Technology, 1200 E. California Blvd., MS 105-24, Pasadena, CA  91125} 
\altaffiltext{2}{Harvard-Smithsonian Center for Astrophysics, 60
  Garden St., MS-66, Cambridge, MA  02138} 


\begin{abstract} 
We present initial results from the M33 Metallicity Project.  Out of
the thousands of cataloged \hii\ regions in M33, only $\sim30$ have
electron-temperature based abundances in the literature.  We have
obtained Keck spectroscopy of a sample of $\sim200$ \hii\ regions in
M33, with 61 detections of the [O~{\sc iii}]~$\lambda 4363$~\AA\ line
that can be used for determining electron temperatures, including
measurements at small galactocentric radii where auroral lines are
generally difficult to detect.  We find an oxygen abundance gradient
of $-0.027 \pm 0.012$~dex~kpc$^{-1}$, in agreement with infrared
measurements of the neon abundance gradient but much shallower than
most previous oxygen gradient measurements.  There is substantial
intrinsic scatter of 0.11 dex in the metallicity at any given radius
in M33, which imposes a fundamental limit on the accuracy of gradient
measurements that rely on small samples of objects.  Finally, we
present a two-dimensional map of oxygen abundances across the southern
half of M33 and discuss the evidence for deviations from axisymmetry.
\end{abstract}


\section{Introduction}

Despite decades of observations of the chemical abundances of stars,
\hii\ regions, and planetary nebulae, the chemical evolution of
galaxies is still a relatively data-starved subject.  Among disk
galaxies (excluding the Milky Way where our location makes it
difficult to discern the global structure), the best studied objects
are M101 (25 \hii\ region abundances; \citealt*{kennicutt03}) and M33
(32 \hii\ region abundances;
\citealt{vilchez88,crockett06,magrini07}).  In both of these galaxies,
the existing metallicity measurements have a spatial sampling rate of
$\la 0.2$~kpc$^{-2}$.  Since \hii\ regions and supernova explosions
typically affect volumes that are less than a few hundred pc in
radius, these data are too sparsely sampled to reveal how chemical
enrichment has proceeded.  In particular, the mixing of metals through
the interstellar medium (ISM) is not yet very well constrained
observationally \citep[e.g.,][and references therein]{se04}, and
whether the spatial distribution of heavy elements is more complex
than the typically assumed simple exponential gradient with radius is
completely unknown.  An additional problem in the study of
extragalactic abundances is the systematic differences between the two
main techniques for measuring abundances: the so-called ``direct''
methods in which the electron temperature is measured directly using
auroral line fluxes, and the empirically calibrated strong-line
methods that use the flux ratios of bright emission lines as a proxy
for the oxygen abundance.  Motivated by these considerations, we have
undertaken the M33 Metallicity Project --- a long term effort to
obtain Keck/LRIS spectroscopy of a sample of $\ga500$ \hii\ regions in
M33 that will provide an unprecedentedly large database of uniformly
determined oxygen, nitrogen, and sulfur abundances in an external
galaxy.

\section{Data and Results}

Using the LRIS spectrograph \citep{oke95} on the Keck~I telescope, we
have observed over 200 \hii\ regions in M33.  Details of the
observations, data reduction, and analysis are given in
\citet[][hereafter RS07]{rs07}.  We detected the [O~{\sc
    iii}]~$\lambda 4363$~\AA\ line and therefore derived direct oxygen
abundances in 61 of the \hii\ regions, approximately tripling the
sample of M33 metallicity measurements available in the literature.

We display our new abundance measurements as a function of radius in
Figure \ref{results}{\emph{a}.  A weak gradient is visible, but the
  intrinsic scatter in the data (which is significantly larger than
  the observational uncertainties) dominates the plot.  We therefore
  must use the method described by \citet{ab96} to determine the
  gradient.  We measure an oxygen abundance gradient of $-0.027 \pm
  0.012$~dex~kpc$^{-1}$, shallower than all previous gradient
  measurements except that of \citet[][hereafter C06]{crockett06}, and
  the intrinsic scatter in the metallicities at any given radius is
  0.11~dex (RS07).

\begin{figure}[!t]
\plottwo{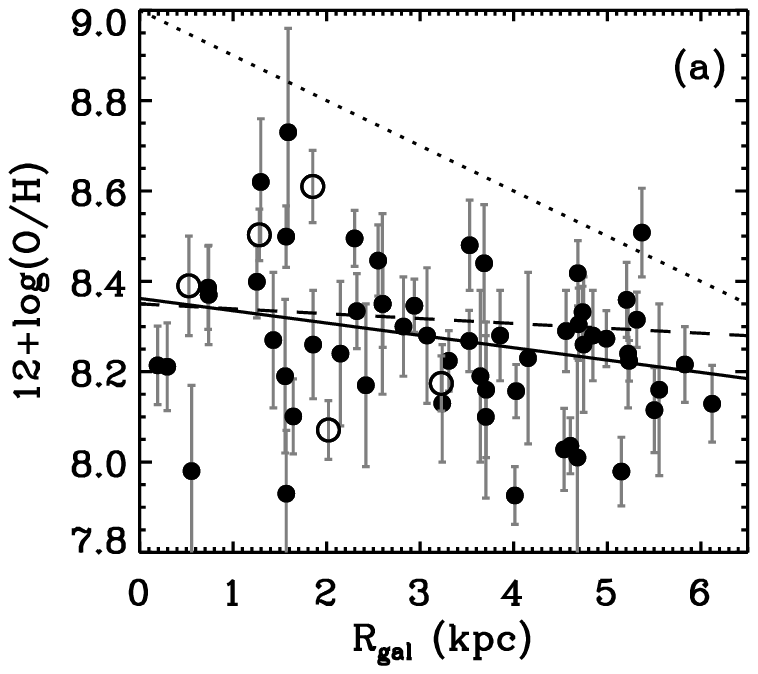}{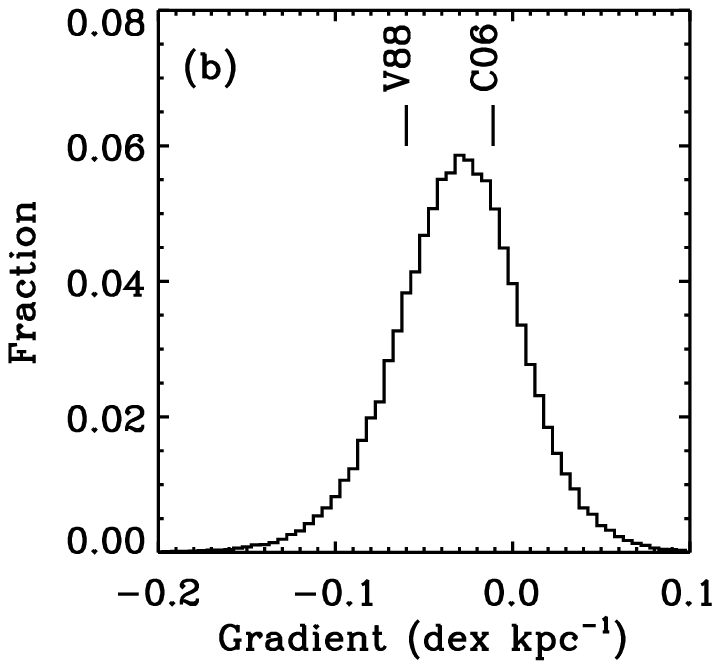}
\caption{(\emph{a}) Oxygen abundance gradient in M33.  A linear
  gradient with a slope of $-0.027$~dex~kpc$^{-1}$ is fit to the data
  (solid line).  Regions with significant He~{\sc ii}~$\lambda
  4686$~\AA\ emission are indicated with open symbols.  The dashed and
  dotted lines represent the gradients measured by \citet{crockett06}
  and \citet{vilchez88}, respectively.  The data are not consistent
  with the steep slopes of $\sim-0.1$~dex~kpc$^{-1}$ found by the
  original M33 studies.  (\emph{b}) Monte Carlo simulation showing the
  distribution of the abundance gradients that would be measured for a
  sample of 10 \hii\ regions drawn randomly from our sample of 61.
  The large width of distribution shows the uncertainty imprinted upon
  the abundance determinations by the underlying variance.  The
  measurements of \citet{vilchez88} and \citet{crockett06} are
  indicated and are consistent with the distribution given their small
  sample sizes. }
\label{results}
\end{figure}

\section{Discussion}

The oxygen abundance gradient that we measure is in good agreement
with infrared measurements of the neon abundance gradient by
\citet{wnp02}, who found a slope of $-0.034 \pm 0.015$~dex~kpc$^{-1}$.
However, our new determination does not appear consistent with most
previous optical studies of the abundance gradient in M33, which
generally found steeper slopes.  Why does our gradient measurement
disagree with these other studies?  For the objects in our sample that
have abundance measurements in the literature, our oxygen abundances
agree within the uncertainties (see RS07), indicating that systematic
errors and sample selection biases are probably not to blame.  Since
we primarily observed \hii\ regions in the southern half of the
galaxy, and most previous studies focused on the northern side, it is
possible that a highly asymmetric abundance distribution could account
for the discrepant results, but this hypothesis can only be tested
with additional observations.

We are therefore left with two possible explanations.  The first is
that \citet[][hereafter V88]{vilchez88}, the largest of the earlier
studies, relied on photoionization modeling for the abundance of one
\hii\ region near the center of M33.  The high derived abundance for
this object significantly steepened the overall gradient they derived
to $-0.10$~dex~kpc$^{-1}$ (after rescaling to our assumed distance);
as V88 noted, the gradient of the outer \hii\ regions in their sample
is only $-0.05 \pm 0.01$~dex~kpc$^{-1}$, much closer to more recent
measurements.  This difference suggests that the photoionization model
abundances may be systematically higher than direct abundances.
Second, the significant intrinsic scatter in the abundances revealed
by our larger sample prevents measurements based on small \hii\ region
samples from accurately determining the gradient.  We plot in Figure
\ref{results}{\emph{b} the outcome of a Monte Carlo simulation in
  which we measured the abundance gradient in M33 by randomly
  selecting 10 \hii\ regions out of our full sample of 61 many times.
  The broad width of the resulting distribution demonstrates that
  significantly larger samples are needed in order to derive accurate
  results.  Even with 61 oxygen abundances, we only detect the
  gradient at $2.3~\sigma$ significance!  Although our measured
  gradient is formally outside the $1~\sigma$ uncertainties of the
  gradient slopes reported by V88, C06, and \citet{magrini07}, when
  the intrinsic scatter and small sample sizes are taken into account
  their results are consistent with our data (see Figure
  \ref{results}{\emph{b}).

In Figure \ref{map}, we display the two-dimensional metallicity
distribution we have derived for M33 overlaid on an H$\alpha$ image of
the galaxy.  As can also be seen in Figure \ref{results}\emph{a}, the
highest metallicity \hii\ regions are not located at the center of the
galaxy, but rather lie along the southern spiral arm at a radius of
$1-2$~kpc.  This distribution suggests that the material enriched by
the most recent generation of star formation in the arm has not yet
been azimuthally mixed through the galaxy.  These data represent some
of the first evidence for a non-axisymmetric abundance distribution in
the interstellar medium of M33.  As we build up a larger data set and
cover the northern side of the galaxy, these measurements will allow
us to constrain the mixing timescale and better understand the
processes by which heavy elements are transported through the galaxy.

\begin{figure}[t]
\plotone{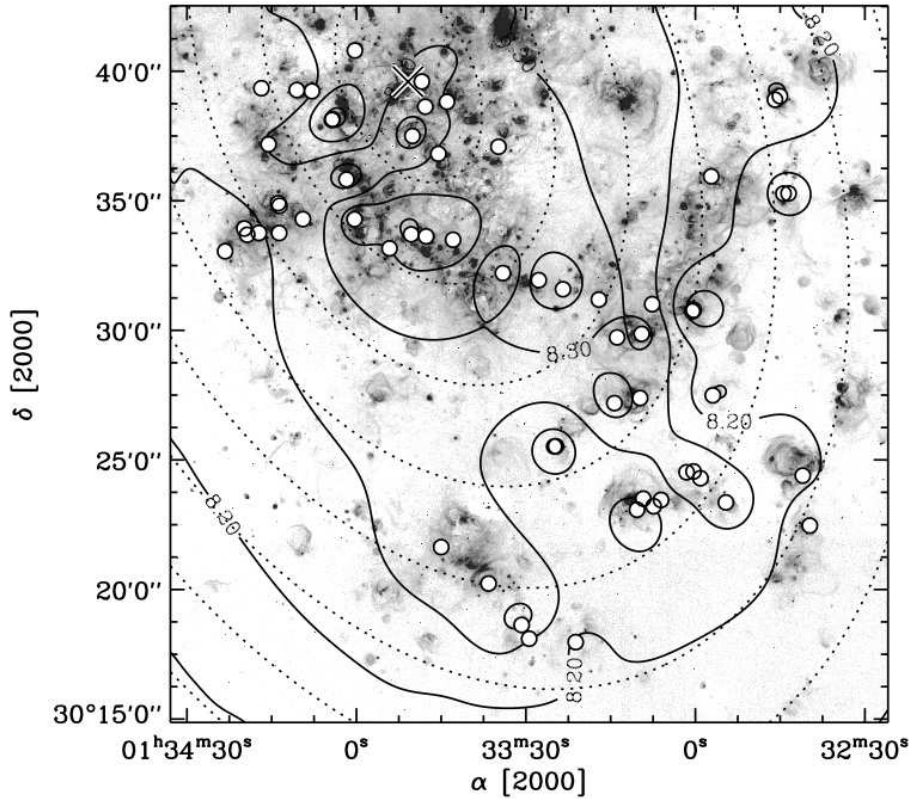}
\caption{Metallicity map of the southwest half of M33.  The white
  circles show the locations of our oxygen abundance measurements, and
  the solid black contours represent metallicity.  The dotted ellipses
  are curves of constant galactocentric radius (spaced at 1~kpc
  intervals), and the center of the galaxy is marked in the upper left
  with an ``X''.  The highest metallicities occur not at the center of
  M33, but in the southern spiral arm.}
\label{map}
\end{figure}

\acknowledgements

JDS acknowledges the support of a Millikan Fellowship provided by
Caltech, and ER acknowledges support from an NSF AAP Fellowship
(AST-0502605).  


\end{document}